\documentclass[aps,prd,floatfix,showpacs,showkeys]{revtex4}

\usepackage{tabularx}
\usepackage{bm}
\usepackage{euscript}
\usepackage{epsfig,psfrag,subfigure,subeqnarray}
\usepackage{graphicx}
\usepackage{color}
\usepackage{amsfonts}
\usepackage{exscale}
\usepackage{amsmath}
\usepackage{amsbsy}
\usepackage{dcolumn}
\usepackage{slashed}

\begin{document}

\title{Another look at charge fractionalization at finite temperature}
\author{Yeong-Chuan Kao and Ming-Chiun Wu}

\affiliation{ Department of Physics, National Taiwan University, Taipei, Taiwan}

\begin{abstract}
We study again the phenomenon of charge fractionalization at finite temperature.  Our calculations are done in a framework in which the connection between the induced fermion number and the chiral anomaly is manifest. We find that the fractional fermion number induced on a soliton
decreases as temperature rises and vanishes at infinite temperature at one-loop level. These results are consistent with previous studies. As an application 
of our approach, we have also studied the behavior of the induced Chern-Simons term in  the (2+1)-dimensional QED at finite temperature.
\end{abstract}

\pacs{11.10.Wx, 11.10.Kk, 12.39.Fe}

\maketitle
\section{INTRODUCTION}

Two of the most intriguing aspects of quantum field theory are the chiral anomaly, which refers to the 
the quantum mechanical breaking of the chiral symmetry\cite{Adler, Bell}, and the charge fractionalization, which refers to the 
fact that fractional fermion numbers can be induced by topologically nontrivial background configurations such as kinks or solitons\cite{Jackiw1976, Su, Goldstone, Bardeen}.  
Since both the chiral anomaly and the charge fractionalization can be understood 
from the viewpoint  of Dirac's negative energy sea\cite{Jackiw1999},
it is perhaps not surprising that a tighter connection  between these two phenomena can be made\cite{Zee, Schaposnik, Mignaco}.
For example, Schaposnik\cite{Schaposnik} showed that, in a commonly used 1+1-dimensional model, the fermion current  receives a contribution from the 
nontrivial Jacobian factor associated with
a chiral rotation of the fermionic field. This contribution leads directly to charge fractionalization.  
Schaposnik's analysis was inspired by Fujikawa's observation\cite{Fujikawa1979, FujikawaSuzuki, Roskies, Saravi} that  the chiral anomaly owes its existence to the 
nontrivial Jacobian associated with the chiral transformation in the path integral. 
Therefore, the nontrivial chiral Jacobian links together the chiral anomaly and the fractional fermion number.
\par 
This connection between the chiral anomaly and the induced fractional fermion number raise the
question: How can the chiral Jacobian and hence the chiral anomaly be temperature independent\cite{Dolan, Itoyama, Reuter, Ruiz, Das1987, Liu, Baier1991, GomezNicola1994, Das1999} while the fractional fermion number is affected by temperature\cite{Niemi, Midorikawa, Aitchison, Dunne}?
This question is very much like the problem of how the amplitude for $\pi^0\rightarrow\gamma\gamma$, which is deeply related to the chiral anomaly, is affected by temperature in the face of the temperature 
independence of the coefficient of the chiral anomaly\cite{Contreras, GomezNicola1993, AlvarezEstrada, Pisarski1996, Pisarski1997, Gelis1999, Gelis2000}.
Here, we present a study of this question. 
We show, in the field theoretic model studied by Shaposnik, how the fractional fermion number acquires a negative temperature-dependent contribution, which 
eventually cancels exactly  the anomaly contribution at infinite temperature at one-loop order. We have been aided by a 
previous study\cite{Kao1998} of the massive Schwinger model in obtaining this result.
The results presented here are consistent with previous studies on induced fermion numbers at finite temperature\cite{Midorikawa, Aitchison, Dunne}.
Our aim is to show that although 
the chiral anomaly  and the fractional fermion number share a common origin they are ultimately very different due to 
their different finite temperature behaviors. 
\par
With slight modifications, our analysis can be immediately applied to the study of the induced Chern-Simons term in 2+1-dimensional QED at finite temperature\cite{Niemi1986, Babu, Poppitz, Aitchison1993, Kao1993, FoscoPRL, FoscoPRD, Brandt}. The reason is that the induced Chern-Simons term is related to the chiral anomaly in a way very similar to the relation between the induced fermion number and the chiral anomaly\cite{FoscoPRL, Niemi1983}. We will show in our approach that the coefficient of induced Chern-Simons term vanishes in the so-called long wavelength limit at infinite temperature, consistent with previous studies.

\section{INDUCED FERMION CURRENT AT FINITE TEMPERATURE}

\par
We  first study the fermionic model  defined by the  following action in two-dimensional (2D) Euclidean space\cite{Jackiw1976, Su, Goldstone, Bardeen, Schaposnik},
	
	\begin{eqnarray}
	 \mathcal{S} = \int_0^{\beta} d\tau \int dx \left[\bar{\psi} i \gamma^{\mu}  \partial_{\mu}\psi + g \bar{\psi} e^{  -\gamma_5 \theta}  \psi \right], \label{S}
	\end{eqnarray}
where $\beta$ is the inverse temperature 1/T, and the Euclidean Dirac matrices $ \gamma^{\mu}$ obey the relations: $	\left\{ \gamma^{\mu}, \gamma^{\nu}\right\}=-2 \delta^{\mu \nu} = 2\eta^{\mu\nu} $, $\mu, \nu = 0 \ \text{or} \ 1$, $\gamma^0=i \sigma_1$, $\gamma^1= i \sigma_2$, $\gamma_5=\sigma_3$, $\gamma^{\mu}\gamma_5= i \epsilon^{\nu \mu} \gamma^{\nu}$, $\epsilon^{01}=-\epsilon^{10}=1$,  $\gamma_{\nu}=-\gamma^{\nu}$, and $\textrm{tr}\gamma_5 \gamma^{\mu} \gamma^{\nu} =-2i \epsilon^{\mu \nu}$. The field $\theta$ represents 
the background soliton. To study the behavior of the fermionic current $\bar{\psi} \gamma^{\mu} \psi $ in the presence 
of $\theta$, we define the following generating functional: 	
	\begin{eqnarray}
	 Z[s]=\int \mathcal{D} \bar{\psi} \mathcal{D} \psi e^{\int_{0}^{\beta} d\tau \int dx  \left[ \bar{\psi} i \gamma^{\mu}  \partial_{\mu}\psi + 
	\bar{\psi}  \gamma^{\mu}  s_{\mu}\psi + g \bar{\psi} e^{ - \gamma_5 \theta}  \psi \right]}, \label{Z}
	\end{eqnarray}
where a source term $s_{\mu}$ is introduced and the expectation value of the fermion current is given by

	\begin{eqnarray}
	J^{\mu}= \langle \bar{\psi} \gamma^{\mu} \psi \rangle =  \frac{1}{Z} \frac{\delta Z}{ \delta s_{\mu} (x)} \bigg|_{s_{\mu}=0}. \label{J}
	\end{eqnarray}
	\\
	\par
	Following Schaposnik\cite{Schaposnik}, we now perform a chiral rotation $\psi= e^{  \gamma_5 \frac{\theta}{2}t} \chi$ ($t$ is a parameter varying from 0 to 1)
and turn the Lagrangian into 
	\begin{eqnarray}
  \mathcal{L}=\bar{\chi} D_t \chi, 
  \end{eqnarray}  
  where 
  \begin{eqnarray}
   D_t \equiv i \gamma^{\mu} \partial_{\mu} + \gamma^{\mu}s_{\mu} +t\gamma^{\mu}a_{\mu}   + g e^{  \gamma_5 \theta (t-1)},
  \end{eqnarray}
  with $a_{\mu}= -\frac{1}{2} \epsilon^{\mu \nu}\partial_{\nu} \theta$.  The point of performing the chiral rotation is to eliminate the $\gamma_5$ exponential in (\ref{S}) for $t=1$ 
  
  Again, following Schaposnik(and Fujikawa), we calculate the Jacobian $\mathcal{J}(\theta) $ associated with the above chiral rotation for $t=1$ at finite temperature 
  
  \begin{eqnarray}
  \mathcal{D} \bar{\psi} \mathcal{D} \psi= \mathcal{J}(\theta) \mathcal{D} \bar{\chi} \mathcal{D} \chi, 
  \end{eqnarray}
  
  \begin{eqnarray}
  \ln \mathcal{J}(\theta)  
  && =- 2\lim_{M \to \infty} \int_0^1 dt \text{Tr} \langle x,\tau | \gamma_5  \frac{\theta}{2} e^{D^2_t/M^2} |x,\tau \rangle \\
  &&= - \lim_{M \to \infty} \int_0^1 dt \ \textrm{tr} \int_{0}^{\beta} d\tau \int dx   \gamma_5  \theta \int_{-\infty}^{\infty} \frac{d k}{2\pi} \frac{1}{\beta}\sum_{\omega_n}e^{-ikx-i\omega_n \tau} e^{D^2_t/M^2} e^{ikx+i\omega_n \tau} \\
  &&= -  \lim_{M \to \infty} \int_0^1 dt \ \textrm{tr} \int_{0}^{\beta} d\tau \int dx   \gamma_5  \theta  \exp\left\{ \frac{i \gamma^{\mu} \gamma^{\nu} (\partial_{\mu}s_{\nu})
                        +it  \gamma^{\mu} \gamma^{\nu} \partial_{\mu}a_{\nu} 
                        +g^2 \gamma_5 \sinh (2\theta(t-1))}{M^2}\right\}  \nonumber \\
  &&  \quad \quad \quad \quad \times \int_{-\infty}^{\infty} \frac{d k}{2\pi} e^{\frac{-k^2}{M^2}} \frac{1}{\beta}\sum_{n=-\infty}^{\infty} e^{\frac{-1}{M^2}\left(\frac{(2n+1)\pi}{ \beta}\right)^2} \label{lnJ9} \\
  &&= - \lim_{M \to \infty} \int_0^1 dt \ \textrm{tr} \int_{0}^{\beta} d\tau \int dx   \gamma_5  \theta  \exp \left\{\frac{i  \gamma^{\mu} \gamma^{\nu} (\partial_{\mu}s_{\nu})
                        +it  \gamma^{\mu} \gamma^{\nu} \partial_{\mu}a_{\nu} 
                        +g^2 \gamma_5 \sinh (2\theta(t-1))}{M^2}\right\} \nonumber \\ 
  && \quad \quad \quad \quad \times \int_{-\infty}^{\infty} \frac{d k}{2\pi} e^{\frac{-k^2}{M^2}} \int_{-\infty}^{\infty} \frac{dk'}{2\pi}  e^{\frac{-k'^2}{M^2}} \label{lnJ10}\\
  &&= - \frac{1}{4\pi} \int_0^1 dt \int_{0}^{\beta} d\tau \int dx    \ \textrm{tr} \left\{\gamma_5  \theta  \left[ i  \gamma^{\mu} \gamma^{\nu} (\partial_{\mu}s_{\nu})
                        +it  \gamma^{\mu} \gamma^{\nu} \partial_{\mu}a_{\nu} 
                        +g^2 \gamma_5 \sinh (2\theta(t-1)) \right] \right\}  \\
  &&= \frac{1}{2\pi} \int_{0}^{\beta} d\tau \int dx  \left[ - \epsilon^{\nu \mu} s_{\nu} \partial_{\mu} \theta
                        + \frac{1}{4} (\partial_{\mu} \theta )^2
                        + \frac{1}{2} g^2 (\cosh2\theta -1) \right] \label{lnJ12},
  \end{eqnarray}
  where in Eq. (\ref{lnJ9}) we have kept only terms which are not zero  in $D_t^2$ after taking the Dirac trace; in Eq. (\ref{lnJ10}) we turn the sum over Matusbara modes 
  into an integral \cite{Ford},
  
  \begin{eqnarray}
  \frac{2\pi}{\beta}\sum_{n=-\infty}^{\infty} e^{-(\frac{(2n+1)\pi}{M\beta})^2} = \int_{-\infty}^{\infty} dk \ e^{-\frac{k^2}{M^2}}.
  \end{eqnarray}  
  In the end,  we have
  
  \begin{eqnarray}
  Z=\exp\left[\frac{1}{2\pi} \int_{0}^{\beta} d\tau \int dx  [ - \epsilon^{\nu \mu} s_{\nu} \partial_{\mu} \theta
                        + \frac{1}{4} (\partial_{\mu} \theta )^2
                        + \frac{1}{2} g^2 (\cosh2\theta -1) ]\right] \int \mathcal{D} \bar{\chi} \mathcal{D} \chi e^{\int_{0}^{\beta} d\tau \int dx   \bar{\chi} \left(i\gamma^{\mu}\partial_{\mu} + \gamma^{\mu} s_{\mu} + \gamma^{\mu} a_{\mu} + g \right) \chi}. \label{Z_result}
  \end{eqnarray}
  \\

  Substituting Eq. (\ref{Z_result}) into Eq. (\ref{J}), we obtain the induced fermion current
  
  \begin{eqnarray}
  J^{\mu}= -\frac{1}{2\pi}\epsilon^{\mu \nu} \partial_{\nu}\theta + j^{\mu}=\frac{a_{\mu}}{\pi}+ j^{\mu},  \label{soliton_current}
  \end{eqnarray}
  where 
  \begin{eqnarray}
  j^{\mu}=\frac{\delta}{\delta a_{\mu}} \ln \det (i \gamma^{\mu}\partial_{\mu} + \gamma^{\mu} a_{\mu} + g ), \label{MSM_current}
  \end{eqnarray}
  is the induced current in the massive Schwinger model\cite{Coleman} with fermion mass $g$ at finite temperature.  Just like the temperature independence of the 
  chiral anomaly\cite{Dolan, Itoyama, Reuter, Ruiz, Das1987, Liu, Baier1991, GomezNicola1994, Das1999}, the coefficient of  the $\frac{a_{\mu}}{\pi}$ term in Eq. (\ref{soliton_current}) is also independent of temperature as expected. Because of the $\frac{a_{\mu}}{\pi}$  term, 
  when we integrate $J^{0}$  over $x$, we find that 
  the induced fermion number is dictated by by the asymptotic values of $\theta$ at $x=\pm \infty$, and hence the fermion number can be fractional in the presence of a soliton.
  The current $ j^{\mu}$ in the massive Schwinger model will only contain derivative terms of  $a_{\mu}$ 
  and therefore 
  does not contribute to the induced fermion number. But  $ j^{\mu}$ will play an important role at finite temperature as is shown in the next section.

\section{FINITE TEMPERATURE EFFECT ON INDUCED FERMION NUMBER}

 \par 
 To see how the induced fermion number obtained in the previous section changes with temperature, we need to calculate 
  vacuum polarization tensor of the massive Schwinger model at finite temperature. 
To one loop-order,
the vacuum polarization tensor at finite temperature is given by
\begin{eqnarray}
 \Pi^{\mu\nu}(p_0,p_1)= \frac{1}{\beta} \sum_n \int_{-\infty}^{\infty} \frac{dk_1}{2\pi} \textrm{tr} \left[ \gamma^{\mu} 
    S_{\beta}(p+k)\gamma^{\nu} S_{\beta}(k) \right], 
\end{eqnarray}
which can be written as\cite{Kao1998}
\begin{eqnarray}
 \Pi^{\mu\nu}(p_0,p_1)=\left(p^2\eta^{\mu\nu}-p^{\mu}p^{\nu}\right) \Pi(p_0,p_1), 
\end{eqnarray}
with
\begin{eqnarray}
 \Pi(p_0,p_1)= \Pi^0(p_0,p_1) + \delta \Pi(p_0,p_1), 
\end{eqnarray}
where $\Pi^0$ is the zero-temperature part, and $\delta \Pi$ is the temperature-dependent part.  As there is a fermion mass gap, the zero-temperature part $\Pi^0$ does not contain a $1/p^{2}$-pole term and therefore could not contribute  to the $a_{\mu}$ term as mentioned above. 
If we want to know how  $ j^{\mu}$ contributes to the fermion number at finite temperature,  we only need to look at (the real part of) $\delta \Pi$. Borrowing the result from \cite{Kao1998}, we know that

\begin{eqnarray}
 \delta \Pi (p_0, p_1) =&&  \frac{2 g^2}{ -p_0^2 - p_1^2} \int_{-\infty}^{\infty} \frac{dk_1}{2\pi} 
\bigg \{ \frac{n_F(E_+)}{E_+} \left[ \frac{1}{(E_+ - p_0)^2- E_-^2} + \frac{1}{(E_+ + p_0)^2- E_-^2}\right] \nonumber \\
&& \quad \quad \quad  \quad \quad \quad \quad \quad + \frac{n_F(E_-)}{E_-} \left[ \frac{1}{(E_- - p_0)^2- E_+^2} + \frac{1}{(E_- + p_0)^2- E_+^2}\right] \bigg\} \label{dPi},
\end{eqnarray}
with $E_{+} = \sqrt{(k_1 + p_1/2)^2+g^2}$ , and $E_{-}=\sqrt{(k_1-p_1/2)^2+g^2}$.

We write $\delta\Pi$ as
\begin{eqnarray}
 \delta \Pi(p_0, p_1) = \frac{1 }{-p_0^2-p_1^2} I(p_0, p_1; \beta g) =  \frac{1 }{p^2} I(p_0, p_1; \beta g), \label{Re_pi}
\end{eqnarray}
where $I(p_0, p_1; \beta g)$ is the integral in Eq. (\ref{dPi}). Straightforward calculations give us the following expression for the static limit of $I(p_0, p_1; \beta g)$:

\begin{eqnarray}
I(p_0=0, p_1 \to 0; \beta g) =\int_{-\infty}^{\infty} \frac{dy}{2\pi} \left\{  \frac{- 2}{ \left( e^{\beta g \sqrt{y^2 + 1}} + 1 \right) (y^2 + 1)^{\frac{3}{2}}} +  \frac{-2 \beta g e^{ \beta g \sqrt{y^2 + 1 }} }{\left(e^{ \beta g \sqrt{y^2 + 1 }}+1\right)^2 (y^2 + 1)}\right\}. \label{I}
\end{eqnarray}
\\
If $I(p_0=0, p_1 \to 0; \beta g)$ is not zero,  $\delta \Pi$ will contain a $1/p^{2}$-pole term and $ j^{0}$ will contain a term proportional to 
$a_{0}$ which can contribute to the induced fermion number when integrated over $x$. If we only keep terms proportional to $a_{0}$, 
the fermion number density $J^{0}$ becomes

\begin{eqnarray}
J^0=\frac{a_0}{\pi} + j^0 = \frac{a_0}{\pi} +\Pi^{00} a_0=\frac{c(\beta g)}{\pi} a_0,
\end{eqnarray}
where
\begin{eqnarray}
c(\beta g)=1+  \pi I(p_0=0, p_1 \to 0; \beta g).
\end{eqnarray}
Numerical integrations allow us to plot $c(\beta g)$ in Fig. 1. We can see that $c(\beta g)$ decreases from 1 at zero temperature to zero at infinite temperature. Therefore, the induced fermion number (obtained by integrating $J^{0}$ over $x$) 
will decrease when temperature increases and vanish completely at infinite temperature.

In comparison with earlier results, we give the following low- and high-temperature limits of Eq. (\ref{I}),

\begin{eqnarray}
 I(p_0=0, p_1 \to 0; \beta g) \approx \left\{ \begin{array}{ll} 
                     - 2\sqrt{\frac{\beta g}{2\pi }} e^{-\beta g}+ \cdots & \text{, $\beta g \gg 1$}\\
                     - \frac{1}{\pi}   + 2 c_2 \beta^2 g^2 + O((\beta g)^3)  & \text{, $\beta g \ll 1$}
\end{array} \right.. \label{Ip00}
\end{eqnarray}
with
\begin{eqnarray}
c_2=\frac{1}{\pi^3} \sum_{n=0}^{\infty} \frac{1}{ (2n+1)^3 }.
\end{eqnarray}

Given the above results, the relevant  terms in $j^{0}$ will be
  
 \begin{eqnarray}
  j^{0}=   I(p_0=0, p_1 \to 0; \beta g)  a_{0}\sim \left\{ \begin{array}{ll} 
                     \left( - 2 \sqrt{\frac{\beta g }{2 \pi  }}  e^{-\beta g}  + \cdots \right) a_{0} & \text{, $\beta g \gg 1$}\\
                     \left( -\frac{1}{\pi} +  2 c_2 \beta^2 g^2 + \cdots \right)a_{0}  & \text{, $\beta g \ll 1$}
\end{array} \right..
  \end{eqnarray}
 and  
  \begin{eqnarray}
c(\beta g) \sim \left\{ \begin{array}{ll} 
                      1 -  \sqrt{2\pi \beta g  }  e^{-\beta g} + \cdots    & \text{, $\beta g \gg 1$}\\
                      2\pi c_2  \beta^2 g^2+ \cdots   & \text{, $\beta g \ll 1$}
\end{array} \right..
\end{eqnarray}
  \\
  These results are consistent with previous studies\cite{Midorikawa, Aitchison, Dunne}.
 
\section{INDUCED CHERN-SIMONS TERM AT FINITE TEMPERATURE}

As an immediate extension of the above analysis, in this section we shall recalculate the much studied coefficient of the induced Chern-Simons term in the (2+1)-dimensional QED at finite temperature\cite{Niemi1986, Babu, Poppitz, Aitchison1993, Kao1993, FoscoPRL, FoscoPRD, Brandt}. It is known that nonanalyticity at $p^{\mu}= 0$ appears in the parity-odd part of the gauge field self-energy $\Pi^{\mu\nu}(p)$ at finite temperature\cite{Kao1993, Brandt}. It turns out that our approach allows the use of a set of background gauge field configurations that will naturally lead us to the long wavelength limit ($p_0\rightarrow 0, \vec{p} = 0$) of the parity-odd part of the self-energy\cite{Kao1993, Aitchison1993, Brandt}, instead of 
the static limit ($p_0 = 0, \vec{p} \rightarrow  0$) commonly obtained in \cite{Niemi1986, Babu, Poppitz, FoscoPRL, FoscoPRD}. Therefore, while the general strategy we employ is the same as that of Fosco, Rossini, and Schaposnik\cite{FoscoPRL, FoscoPRD}, our result also nicely complements  theirs. 

\par
The theory we study is given by the following action in three-dimensional (3D) Euclidean space:

	\begin{eqnarray}
	 \mathcal{S} = \int_0^{\beta} d\tau \int_0^{L_1}  d x_1 \int_0^{L_2}  d x_2 \ \bar{\psi}(\tau,x_1, x_2)\left[ i\slashed \partial +   e\slashed a + M \right] \psi(\tau,x_1,x_2), 
	 \label{Sqed3}
	\end{eqnarray}
with the 3D Euclidean Dirac matrices $\gamma^{\mu}$ chosen to be  $\gamma^0= i \sigma_1,\gamma^1=i\sigma_3 ,\gamma^2=i\sigma_2$. For convenience, we let  $x_1\in[0,L_1]$ and  $x_2\in[0,L_2]$, with  $L_1$ and  $L_2$ serving as large infrared cutoffs. The goal is  to integrate out the fermionic fields to find the effective action as was done in previous sections.  We choose to have a spatially uniform but time-dependent electric field as the background  electromagnetic field. Therefore, the class of gauge field configurations corresponding to this particular kind of electric field can be set as 
$a_{\mu}(\tau)= (a_0(\tau), a_1( \tau), a_2(\tau)) $. For comparisons, we point out that  a time-independent magnetic field is employed in \cite{FoscoPRL, FoscoPRD} as the  background electromagnetic field.

\par

The first step of the analysis is to Fourier transform the fermionic field over the $x_1$ coordinate as

\begin{eqnarray}
&& \psi(\tau, x_1, x_2) = \frac{1}{L_1}\sum_{n} e^{-ik_n x_1} \psi_{k_n}(\tau, x_2) \quad ;   \quad \text{with} \  k_n=\frac{(2n+1)\pi}{L_1},
\end{eqnarray}
then the action becomes

\begin{eqnarray}
 S = \frac{1}{L_1}\sum_{n} \int_0^{\beta} d\tau \int_0^{L_2} d x_2 \ \bar{\psi}_{k_n}(\tau,x_2)\left[\slashed d +   \gamma^1 (k_n +ea_1) + M\right] \psi_{k_n}(\tau,x_2),
\end{eqnarray}
where
\begin{eqnarray*}
&& \slashed d = \gamma^0 (i \partial_0 + e a_0)+\gamma^2 (i \partial_2 + e a_2).
\end{eqnarray*}
Hence, the problem of obtaining the determinant of the 3D Dirac operator $\text{det}[i \slashed \partial +   e\slashed a + M]$ becomes that of obtaining the determinant of the reduced 2D Dirac operators, which we write as
\begin{eqnarray}
\text{det}[\slashed d +   \gamma^1 (k_n +ea_1) + M] = \text{det}[\slashed d +  \rho_{k_n}(\tau) e^{i\gamma_5 \phi_{k_n}(\tau) } ], \label{QED3det}
\end{eqnarray}
where $\gamma_5 \equiv \sigma_3$
and
\begin{eqnarray}
&& \rho_{k_n}(\tau) = \sqrt{M^2 + (k_n+e a_1(\tau))^2} ; \\
&& \phi_{k_n}(\tau) = \tan^{-1} \left( \frac{k_n+ e a_1(\tau) }{ M } \right). \label{3Dphi}
\end{eqnarray}
Following the strategy explained in section II and in \cite{FoscoPRL, FoscoPRD}, we now make a chiral rotation to transform away the factor $e^{i\gamma_5 \phi_{k_n}}$  in Eq. (\ref{QED3det}) so that 
the determinant becomes
\begin{eqnarray}
&& \text{det}[\slashed d +  \rho_{k_n}(\tau)e^{i\gamma_5 \phi_{k_n}(\tau)} ] = \mathcal{J}_{k_n} \text{det} [\slashed D +  \rho_{k_n}(\tau)],
\end{eqnarray}
where $\slashed D = \gamma^0 (i \partial_0 + e a_0 +  b_0)+ \gamma^2 (i \partial_2 + e a_2 +  b_2)$,  $b_0 = \frac{i}{2}  \partial_2 \phi_{k_n}=0$, $b_2 = -\frac{i}{2}  \partial_0 \phi_{k_n}$. Note that the 2D Dirac operator $\slashed D +  \rho_{k_n}(\tau) $ is that of the massive Schwinger model with fermion mass $\rho_{k_n}(\tau) $ . We easily find the Fujikawa Jacobian $\mathcal{J}_{k_n}$ to be  
\begin{eqnarray}
&& \ln  \mathcal{J}_{k_n}=  -i \frac{e}{2\pi}   L_2 \int_0^{\beta} d\tau  \left[  -\phi_{k_n} \partial_0 a_2 - \frac{i}{4} (\partial_{0}\phi_{k_n} )^2+ \frac{1}{2} \rho_{k_n}^2 (\cos2\phi_{k_n} -1)\right],
\end{eqnarray}
by adapting Eq. (\ref{lnJ12}) to the present case. In the end we have
\begin{eqnarray}
\ln \text{det}[\slashed \partial +  i e\slashed a + M] &=& \sum_{n} \ln \left\{ \mathcal{J}_{k_n} \text{det} [\slashed D  +  \rho_{k_n}(\tau)] \right\}  \\
&\equiv&\ln \mathcal{J} + \sum_{n} \ln \text{det} [\slashed D  +  \rho_{k_n}(\tau)] \label{lndet},
\end{eqnarray}
with
\begin{eqnarray}
\ln  \mathcal{J} &=& \sum_{n} \ln  \mathcal{J}_{k_n}  \label{3DlnJ} \\
&=&  i \frac{e}{2\pi} L_2 \sum_{n} \int_0^{\beta} d\tau  \left[\tan^{-1} \left( \frac{k_n+ ea_1 }{ M } \right)  \partial_0 a_2\right] \\
&=& i\frac{M}{|M|} \frac{e^2  }{4\pi} L_1 L_2 \int_0^{\beta} d\tau   a_1 \partial_0 a_2 \\
&=&  i\frac{1}{2} \frac{M}{|M|} \frac{e^2  }{4\pi} L_1 L_2 \int_0^{\beta} d\tau   \left(a_1 \partial_0 a_2 - a_2 \partial_0 a_1\right). \label{3DlnJresult}
\end{eqnarray}
It is seen that the Fujikawa Jacobian leads to the following induced Chern-Simons term in the effective action in our chosen background field,
\begin{eqnarray}
S_{\text{eff}}&=&  - i\frac{1}{2}\frac{M}{|M|} \frac{e^2  }{4\pi} \int_0^{\beta} d\tau \int_{-\infty}^{\infty}\int_{-\infty}^{\infty}  d^2x  \ \epsilon^{\mu \nu \alpha} \ a_{\mu} \partial_{\nu} a_{\alpha}.
\end{eqnarray}
\par
Next, we consider the contributions from  $\text{ln}\text{det} [\slashed D +  \rho_{k_n}(\tau)] $ in Eq. (\ref{lndet}). At zero temperature, these contributions vanish, but they become nonvanishing at finite temperature similar to the contribution to the induced fermion number from the $j^{\mu}$ term  in Eq. (\ref{MSM_current}) .
The one-loop contribution from $\text{ln}\text{det} [\slashed D +  \rho_{k_n}] $ to the local Chern-Simons term is again easily obtained from the vacuum polarization tensor $\Pi^{i j}(p_0,p_2)=\left(p^2\eta^{i j}-p^{i}p^{j}\right) \Pi(p_0,p_2)$ (with $i,j= 0\ \text{or} \ 2$, and $\eta^{00}=-1=\eta^{22}$) of the massive Schwinger model given in Sec. III in the long wavelength limit $x_2\in[0,L_2]$
\begin{eqnarray}
S_{k_n}
&\equiv&\ln \text{det} [\slashed D  +  \rho_{k_n}(\tau, x_2)] \\
&=& \frac{L_2}{2}\int_0^{\beta} d\tau  (ea_{i}+ b_{i}) \cdot [\Pi^{i j}(p_0 \to 0, p_2 = 0)] \cdot (ea_{j}+ b_{j}) + \cdots \\
&=& \frac{L_2}{2}\int_0^{\beta}  d\tau  I(p_0 \to 0, p_2 = 0;\beta \rho_{k_n}) \eta^{i j} (ea_{i}+ b_{i})  (ea_{j}+ b_{j}) + \cdots \label{MSM3D},
\end{eqnarray}
where $I(p_0 , p_2 ;\beta \rho_{k_n}) $ is defined by Eq. (\ref{Re_pi}) with $p_1$ replaced by $p_2$ and $\beta g$ replaced by $\beta \rho_{k_n}$ .

Again by using the result in \cite{Kao1998}, we have the following long wavelength limit
\begin{eqnarray} 
 I(p_0 \to 0, p_2 = 0; \beta \rho_{k_n}) = \int_{-\infty}^{\infty}  \frac{dq}{2\pi} \frac{- 2 } { \left( e^{ \beta \rho_{k_n}\sqrt{q^2 + 1}} + 1 \right) (q^2 + 1)^{\frac{3}{2}}} \label{Ip10}.
\end{eqnarray}
Substituting Eq. (\ref{Ip10}) into Eq. (\ref{MSM3D}),  we see that $S_{k_n}$ contains a term of the form
\begin{eqnarray}
&&  -i \frac{e}{2} L_2 \int_0^{\beta}  d\tau  I(p_0 \to 0, p_2 = 0;\beta \rho_{k_n})  a_2 \partial_0 \phi_{k_n}  \\
&&=  i \frac{M}{|M|}\frac{e}{2} L_2 \int_0^{\beta}  d\tau  \int_{-\infty}^{\infty} \frac{dq}{2\pi} \frac{- 2  \phi_{k_n}(|M|)} { \left( e^{ \beta \rho_{k_n}\sqrt{q^2 + 1}} + 1 \right) (q^2 + 1)^{\frac{3}{2}}} \partial_0 a_2.   
\end{eqnarray}
Hence, the contributions from  $\sum_{n} S_{k_n}$ to the effective action contains the following term 

\begin{eqnarray}
&&i\frac{M}{|M|} \frac{e}{4\pi} L_2 \sum_{n} \int_0^{\beta}  d\tau  \int_{-\infty}^{\infty} dq \frac{- 2  \phi_{k_n}(|M|)} { \left( e^{ \beta \rho_{k_n}\sqrt{q^2 + 1}} + 1 \right) (q^2 + 1)^{\frac{3}{2}}} \partial_0 a_2 \\
&&\to i\frac{M}{|M|} \frac{e}{4\pi} \frac{L_1 L_2}{2\pi} \int_{-\infty}^{\infty} dk \int_0^{\beta}  d\tau  \int_{-\infty}^{\infty} dq \frac{- 2  \phi_{k}(|M|)} { \left( e^{ \beta \rho_{k}\sqrt{q^2 + 1}} + 1 \right) (q^2 + 1)^{\frac{3}{2}}} \partial_0 a_2,  \label{QEDMSM1} 
\end{eqnarray}
where we approximate the sum over $n$ by an integral over $k$ for very large  $L_1 $. Expanding Eq. (\ref{3Dphi}) to first order in $e a_1$, we find
\begin{eqnarray}
\int_{-\infty}^{\infty} d k \frac{- 2  \phi_{k}} { \left( e^{ \beta \rho_{k}\sqrt{q^2 + 1}} + 1 \right) (q^2 + 1)^{\frac{3}{2}}} 
&=& \int_{-\infty}^{\infty} d k \Bigg\{\frac{-2}{\left(e^{\beta |M| \sqrt{(k^2 +1 )(q^2 +1)}}+1\right)(1+k^2) (q^2 + 1)^{\frac{3}{2}}} \nonumber \\
&& \quad \quad \quad \quad + \frac{2 \beta |M| (\tan^{-1}k) k  e^{\beta |M| \sqrt{(k^2 +1 )(q^2 +1) }} }{\left(e^{\beta |M| \sqrt{(k^2 +1 )(q^2 +1)}}+1\right)^2 (q^2 + 1) \sqrt{k^2+1} } \Bigg\}e a_1 + \cdots,
\end{eqnarray}
At the end, we find that in addition to Eq. (\ref{3DlnJ}), the effective action contains an extra Chern-Simons term of the form
\begin{eqnarray}
&& -i c_3(\beta |M|) \frac{M}{|M|} \frac{ e^2}{4\pi} L_1 L_2 \int_0^{\beta}  d\tau   a_1 \partial_0 a_2  \\
&&=  -i \frac{c_3(\beta|M|)}{2} \frac{M}{|M|} \frac{e^2  }{4\pi} \int_0^{\beta} d\tau \int_{-\infty}^{\infty} \int_{-\infty}^{\infty} d^2x \ \left( a_1  \partial_0 a_2 - a_2 \partial_0 a_1  \right), \label{c33}
\end{eqnarray}
where 
\begin{eqnarray}
c_3(\beta|M|) &\equiv& \int_{-\infty}^{\infty} dk  \int_{-\infty}^{\infty} \frac{dq}{2\pi}   \Bigg\{\frac{2}{\left(e^{\beta |M| \sqrt{(k^2 +1 )(q^2 +1)}}+1\right)(1+k^2) (q^2 + 1)^{\frac{3}{2}}} \nonumber\\
&& \quad \quad \quad \quad \quad \quad \quad \quad - \frac{2 \beta |M| (\tan^{-1}k) k  e^{\beta |M| \sqrt{(k^2 +1 )(q^2 +1) }} }{\left(e^{\beta |M| \sqrt{(k^2 +1 )(q^2 +1)}}+1\right)^2 (q^2 + 1) \sqrt{k^2+1} } \Bigg\}.
\end{eqnarray}

Combining Eq. (\ref{3DlnJresult}) and Eq. (\ref{c33}), we finally have the full-induced Chern-Simons term in the effective action at finite temperature,

\begin{eqnarray}
S^{\beta}_{\text{eff}}&=& i \frac{(1-c_3(\beta|M|) )}{2} \frac{M}{|M|} \frac{e^2  }{4\pi} \int_0^{\beta} d\tau \int_{-\infty}^{\infty} \int_{-\infty}^{\infty} d^2x \ ( a_1  \partial_0 a_2 - a_2 \partial_0 a_1) \\
&\equiv&   i \frac{\kappa(\beta|M|)}{2} \frac{M}{|M|} \frac{e^2  }{4\pi} \int_0^{\beta} d\tau \int_{-\infty}^{\infty} \int_{-\infty}^{\infty} d^2x \ \left( a_1  \partial_0 a_2 - a_2 \partial_0 a_1  \right)\\
&=& -i\frac{\kappa(\beta|M|)}{2}\frac{M}{|M|} \frac{e^2  }{4\pi} \int_0^{\beta} d\tau \int_{-\infty}^{\infty}\int_{-\infty}^{\infty}  d^2x  \ \epsilon^{\mu \nu \alpha} \ a_{\mu} \partial_{\nu} a_{\alpha}.
\end{eqnarray}
We obtain the coefficient of the induced Chern-Simons term $\kappa(\beta|M|)$ as a function of $\beta|M|$ by numerical integrations. The result is plotted in Fig.(\ref{fig:c4}). We see that $\kappa(\beta|M|)$ vanishes at infinite temperature in agreement with a previous study in the long wavelength limit\cite{Kao1993}.

\section{DISCUSSIONS}
Schaposnik has shown that the induced fermion number on a soliton can be seen as arising from the Fujikawa jacobian associated with a chiral rotation. 
We have shown in Schaposnik's approach that the induced fractional fermion number on a soliton will indeed reduce smoothly to zero at one-loop order when temperature increases to infinity. (As a byproduct of our analysis, we have also shown that the coefficient of the induced Chern-Simons term in 2+1 QED vanishes in the long wavelength limit at infinite temperature.) This behavior is in contrast with the well-known temperature independence of the chiral anomaly.
Therefore, while the phenomena of charge fractionalization and chiral anomaly  are related to each other, 
they are in the end very different, as seen most clearly at finite temperature. Besides the analogy to the temperature dependence of the 
amplitude for $\pi^0\rightarrow\gamma\gamma$ mentioned in the introduction, we also want to point out that 
the fate of the induced fermion number is not 
unlike that of the chiral condensate $\langle \bar{\psi}\psi \rangle$ in the massless Schwinger model\cite{Lowenstein, Nelson}. It is known that without the chiral anomaly to spoil the chiral symmetry, there will be no nonvanishing $\langle \bar{\psi}\psi \rangle$ in the massless Schwinger model because the Coleman theorem\cite{Coleman1973} forbids spontaneous breakdown of continuous symmetries in 2D. In a way, we may say that $\langle \bar{\psi}\psi \rangle$ owes its existence to the chiral anomaly.  Yet, $\langle \bar{\psi}\psi \rangle$ 
still decreases smoothly to zero at infinite temperature\cite{Kao1992} even if the chiral anomaly is not affected by temperature. 
\par

\newpage
\begin{figure}[t]
\includegraphics[width=0.5\textwidth, clip]{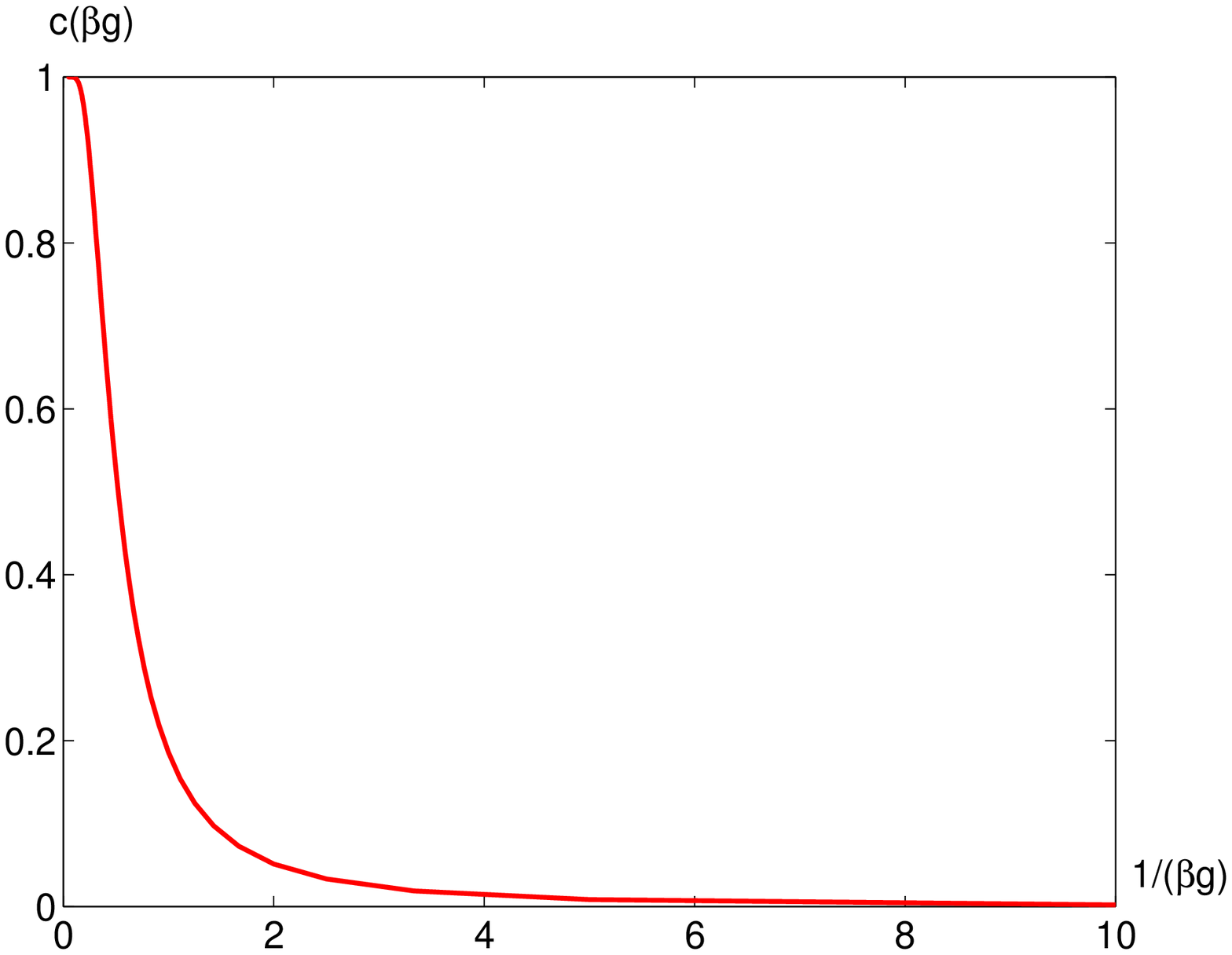}
\caption{{ \protect\small {Plot of $c(\beta g)$ defined by  Eq. (\ref{Ip00})}}}
\label{fig:IT}
\end{figure}
\begin{figure}[t]
\includegraphics[width=0.5\textwidth, clip]{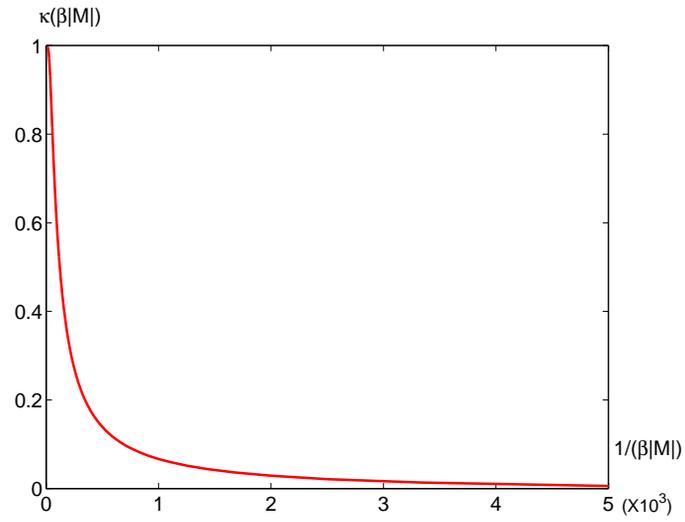}
\caption{{ \protect\small {Plot of $\kappa(\beta |M|)$}}}
\label{fig:c4}
\end{figure}

\end{document}